# Diversity, Equity and Inclusivity in the Australian and New Zealand Medical Physics Workforce

Emily Simpson-Page [1][0000-0002-5262-6465], Megan Anderson[2] Scott Crowe[1,3,4][0000-0001-7028-6452] Tanya Kairn[1,3,4][0000-0002-2136-6138] Ian Smith[5][0000-0001-9784-0855] Christine Thompson[6] Paul Keall[7][0000-0003-4803-6507]

[1] Royal Brisbane & Women's Hospital, Herston, QLD, Australia
[2] Australasian College of Physical Scientists and Engineers in Medicine, Mascot, NSW, Australia
[3]School of Information Technology and Electrical Engineering, University of Queensland, St Lucia, QLD, Australia
[4]School of Chemistry and Physics, Queensland University of Technology, Brisbane, QLD, Australia
[5]St Andrews War Memorial Hospital, Spring Hill, QLD, Australia;
[6]Te Pūriri o Te Ora, Auckland District Health Board, Auckland, New Zealand;
[7]ACRF Image X Institute, University of Sydney, Sydney, Australia
emily.simpson-page@health.qld.gov.au

**Abstract.** The consideration of diversity, equity and inclusivity (DEI) is an important part of promoting a robust and respectful workforce, and is critical to the continued success of organisations, including healthcare providers, academic institutions and professional societies. Many professional bodies representing medical physicists have made commitments to DEI principles in the form of mission statements, policies, steering groups, frameworks and workforce surveys.

In the Australian and New Zealand medical physics community, DEI work has included reflecting on the impact of stereotypes, surveys on workforce experiences, and capturing workforce diversity metrics (including gender, nationality, age and professional background). These projects have been conducted with the support of the Australasian College of Physical Sciences and Engineering in Medicine (ACPSEM) and have contributed to the enhancement of DEI in the ACPSEM workforce. Most of this work has been focused on gender diversity, reflecting increasing involvement in the Australian and New Zealand workforce: women accounted for 41% of medical physics trainees and 32% of registered medical physicists in a 2020 survey.

In 2021, the ACPSEM Professional Standards Board (PSB) incepted the creation of a DEI working group. This movement was supported not only by the ACPSEM Board but also the wider medical physics community, both nationally and internationally. The establishment of this group will support best-practice governance, define the diversity metrics most relevant to our profession and use data collected on these metrics to effect positive change within the ACPSEM boards and committees as well as embedding DEI in policies and procedures for internal and external activities. The DEI working group will enhance the ACPSEM's reputation and ultimately ensure all members feel empowered, respected and represented.

# 1 Introduction

The creation and promotion of Diversity, Equity and Inclusivity (DEI) efforts has been, and is being, adopted by physics communities worldwide. Multiple institutions and governing bodies have committed to goals to improve representation in physics workforces including the Institute of Physics [1], Institute of Physics and Engineering in Medicine [2] (UK), American Association of Physicists in Medicine multiple committees on diversity and inclusion [3] and the International Atomic Energy Agency including diversity in their core ethical values [4]. These efforts are in line with The United Nations fifth sustainable development goal of gender equality and women's empowerment [5], Springer Nature Groups commitment to advancing DEI [6], other initiatives promoting the representation of lesbian, gay, bisexual and transgender (LGBTQ+) communities within physics [7] and ending discrimination towards marginalized groups in healthcare [8]. A key motivation behind these efforts is the knowledge that diverse and inclusive teams "outperform homogeneous groups on complex tasks" and are "better equipped to address health disparities" [9]. Moreover, inclusive work environments have been shown to foster more creativity, productivity and innovation within a workplace [10]. Similarly, by promoting inclusivity and psychological safety, untapped ingenuity adds to the overall state and prosperity of a workforce [11]. In the New Zealand context ensuring equity is not optional but mandated in law (12). Te Tiriti o Waitangi (the Treaty of Waitangi), the founding treaty of New Zealand between the Māori and the British crown requires that Māori are given equal rights to people of European descent. Historically this has not been complied with and all equity work should seek to address historical inequities. These themes of inclusiveness, diversity and respect have also been explored as vital to ensure safety in radiation practice, whereby a team that fosters respect and psychological safety will be more likely to report near miss events in radiation oncology [13].

In Australia and New Zealand, work on the DEI landscape has included reflection on the impact of stereotypes [14], workforce analysis [15,16] and recently, the creation of a DEI Working Group for the Australasian College of Physical Scientists and Engineers in Medicine (ACPSEM). One of the main goals of the working group will be to gain more of an understanding of the minorities represented, and those not represented or underrepresented, within the community and promote an inclusive environment. This body of work summarizes the DEI efforts already undertaken in Australia and New Zealand, highlights some key findings from the most recent workforce survey and summarises the process of the creation of a DEI working group for the ACPSEM.

# 2 Previous DEI efforts in Australia and New Zealand Workforce

In early 2016, a preliminary analysis of the Australian and New Zealand workforce and research participation of women in Medical Physics was completed [16]. This study identified the gap of data available of the gender balance in the medical physics workforce and sought to provide a baseline from which future studies and investigations could compare. This action to collect data and observe trends over time is imperative and was a vital step to observing differences in gender across a cohort [17] and identify marginalized groups. Following this analysis, an editorial was published in the ACPSEM journal, Physical and Engineering Sciences in Medicine, titled ‘The impact of diversity, bias and stereotype: expanding the Medical Physics and Engineering (Science Technology Engineering and Mathematics) STEM workforce’ [14]. This editorial highlighted some of the systemic challenges faced by females and made actionable suggestions such as “Increasing the diversity in keynote speakers at conferences and on professional councils [as] a public demonstration of diversity embraced”. Two years later in 2018, a survey was designed to determine the workplace experiences of male and female members of the ACPSEM to identify potential barriers limiting minorities in the workforce. This survey identified opportunities for mentorship among other strategies to help reduce bias against women and tackle barriers for professional development and career progression [15]. Note that the survey only offered a binary gender choice, female or male.

## 3 Workforce Survey

Workforce surveys and data collection provides useful insights into the landscape of a workforce, as previous efforts have shown. In 2020, a Radiation Oncology Medical Physics (ROMP) workforce survey undertaken through the distribution of two surveys across Australia and New Zealand [18]. The survey captured information of gender in the workforce as well as demographic data and working habits. The workforce modelling project was supported by the ACPSEM and had a main objective to understand workforce requirements.

The survey included 182 individual member respondents and 98 radiation oncology departments. Of these 182 members, 64% identified as male and 36% as female. The ACPSEM registration data base was also used to gain an understanding of the total workforce: 334 registered ROMPs with 69% male and 31% female identifying cohort (noting that the ACPSEM registration website currently only had binary “Male” and ”Female” options, the workforce survey did include a total of six options: "Female", "Male", "Intersex", "Trans", "Non-Conforming" and "Personal").

Although the workforce survey did not have systems set up to capture sufficient data to look at diversity other than gender, age, current location and place of certification, it did provide a good starting point for expected representation on panels and boards. The capture of gender metrics can be compared against a workforce analysis from 2016 [16], which reported an estimate of 28% of the workforce identifying as being female and a 2008 survey [19] which reported 53% of the 20-34 year old cohort identified as female and 33% of all ROMPS (including registrars) identified as female. Based on these surveys, over a fourteen year period the representation of females in the workforce has disappointingly not increased. To understand the percentage of the workforce that is female is a starting point for the expected percentage of female representation on boards and panels [20].

The 2020 workforce survey highlighted some differences between male and female respondents and their work activities and outlined the current landscape of the workforce in terms of residence, certification location, age and career stage. Due to the number of respondents, and the relatively small workforce in Australasia, the results of the survey allowed an understanding of trends, but many of the respondent categories were too small to conclude statistically significant results.

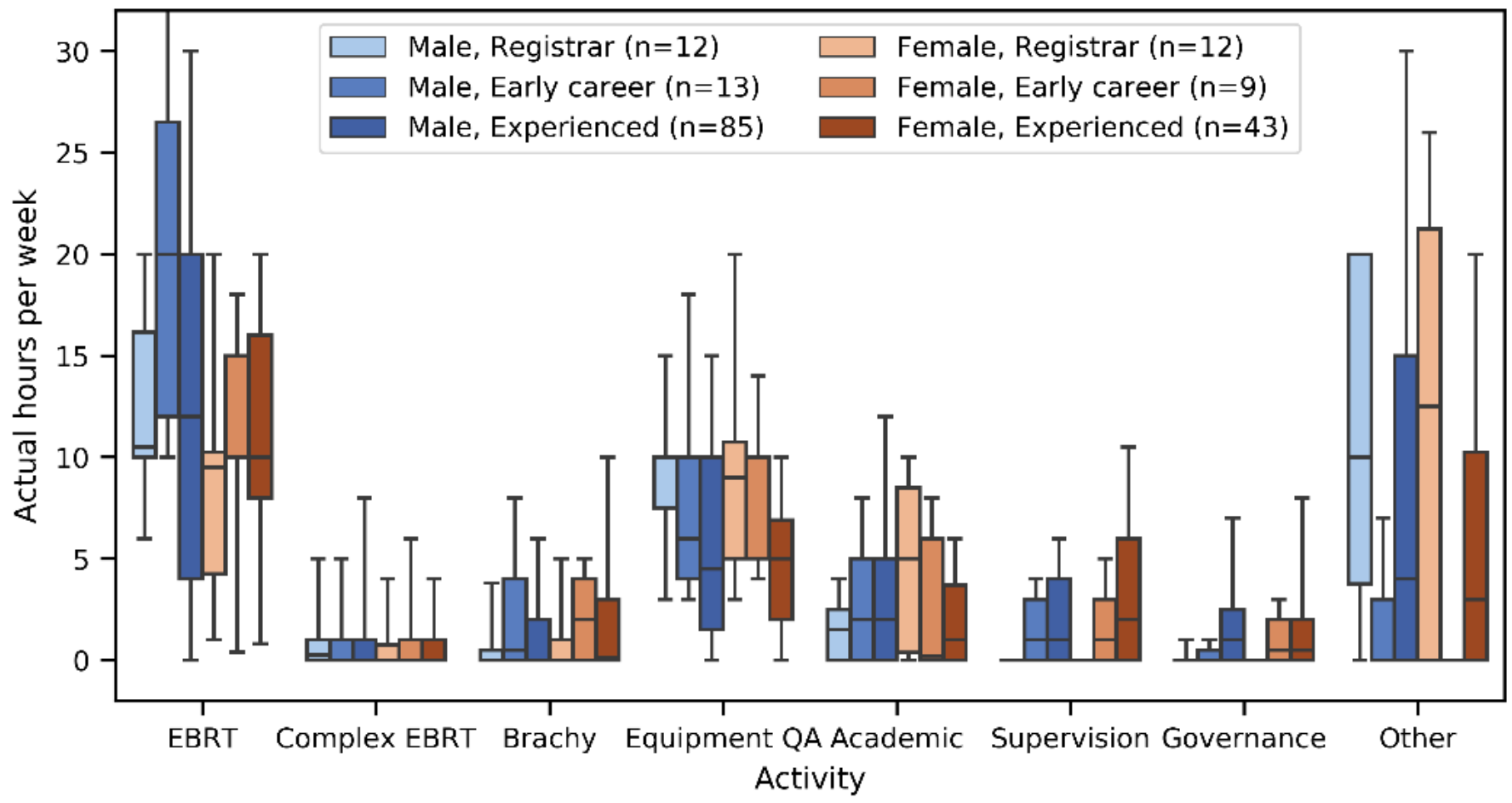

**Fig. 1.** Snapshot of workforce survey with male and female respondents, separated into career stage, for reported actual hours per week. The box in the boxplot indicates median and interquartile range, whiskers indicate 5$^{th}$ and 95$^{th}$ percentiles.

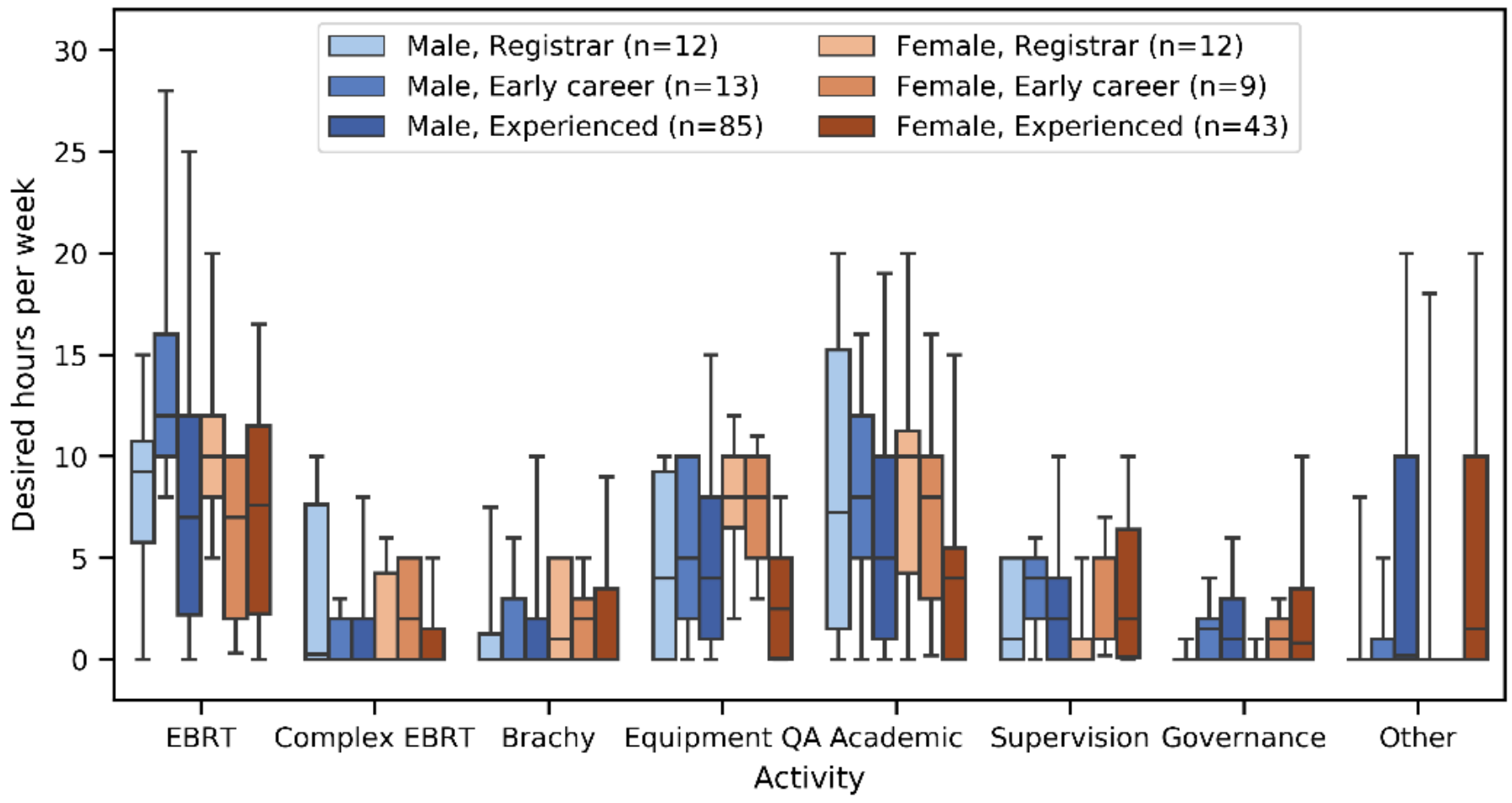

**Fig. 2.** Desired hours per week for male and female respondents, separated into career stage, from the 2020 workforce survey. The box in the boxplot indicates median and interquartile range, whiskers indicate 5$^{th}$ and 95$^{th}$ percentiles.

Overall, the work reported by male and female respondents was similar as was the desired work distribution [Fig 1, Fig 2]. The largest median differences between genders were current and ideal working hours; males reported more external beam quality assurance work, such as plan checking and dose measurements whereas female respondents reported a higher percentage of supervision work. In terms of academic work, for actual hours reported, the median number of hours did not change for male respondents with career progression, whereas for female respondents it trended down with experience. Barriers to women feeling empowered and supported to complete academic work (often in addition to routine clinical work) have been explored previously [15]. Similarly, efforts to increase the international visibility of female representation within the scientific community have worked against the observation of these trends [21]. Both male and female respondents in early career stage reported a desire for more hours in academic and supervision work.

Regarding other demographics, the average age of certified radiation oncology medical physicists was 43.3 years. Of the 181 respondents who indicated where they were living, 81% were in Australia, 18% were in New Zealand, and 1 was living overseas. Of the 151 registered ROMPs that responded, 60% were certified in Australia, 11% were certified in New Zealand, and 13% were certified in the United Kingdom.

## 4 Creation of DEI Working Group

In 2018 the ACPSEM undertook a project to refresh its professional standards which saw recommendations implemented to include representation from different career stages and subspecialties into governance processes. This promotion of representing certain minorities in the profession proved successful and in 2020 the board approved recommendations from the PSB to create a DEI working group. The Terms of Reference for the group included six general objectives:

1. Undertake detailed review of existing processes and policies and make recommendations to Board where potential improvements may exist
2. Develop considered DEI Policy
3. Develop a DEI focused membership survey
4. Consider where DEI efforts should be focused from membership survey
5. Brainstorm how ACPSEM could potentially improve and support DEI initiatives
6. Develop educational materials on DEI work

In January 2022, an Expression of Interest (EOI) for the working group went out to members. The criteria for the group included:

**Essential:**
Participation in DEI brainstorming workshops, as required
Attending meetings via Zoom to define scope of work and allocate workload, as required
Undertake agreed tasks within agreed time frames
Work collaboratively to execute agreed work and provide summaries for ACPSEM Board for approval
Experience and/or interest in driving DEI principles at ACPSEM or elsewhere

**Desirable:**
A passion for social equity and inclusivity with lived experience of diversity
Experience in the implementation of DEI in processes and systems

The working group EOI requested at least five members for the working group and the selection panel consisted of a combination of ACPSEM staff, Board and current or former PSB members. Eight members were included in the final DEI working group, against the outlined selection criteria. This consisted of six females and two males with a variety of lived experiences in diversity and each from different backgrounds including Māori, LGBTQ+, disabled communities, working parents, various career stages and various ages.

Lessons learned from the EOI process, to be implemented for future working group membership applications, included a review of the selection criteria, some more explicit metrics to assess candidates against (such as previous ACPSEM college experience) and clear instructions for ranking candidates. Questions were also raised during the EOI process on the robustness of the selection; however, the importance of understanding the landscape of the workforce and the minorities requiring representation was difficult without further information. During the selection process, the panel discussed inherent bias, and it is planned that the working group will ensure it accounts for minorities and intersectionality, ensuring reduced barriers to entry and promotion of a more equitable workforce, as similarly explored in the United States [22]

To support these goals, a new workforce survey capturing more information including migrant status, ethnicity, gender identity, disability and indigenous Australians and New Zealanders, with an initial focus on educational barriers for Aboriginal, Torres Strait Islanders, Māori and Pacifica, as well as other diversity metrics will be undertaken in 2022. Moreover, the working group will develop a DEI policy for board review and potentially further DEI resources such as the Institute of Physics DEI white paper outlining their five drivers for improving DEI [23]. The working group will also work in coordination with the Engineering & Physical Sciences in Medicine conference to promote a safe, respectful and inclusive conference environment with diverse representation of speakers.

## 5 Conclusion

Ultimately, all of the efforts outlined in this study are working towards a common goal: To promote under-represented groups in the Australia and New Zealand workforce, to foster a safe, diverse and inclusive workplace for everyone and to ensure equal opportunities for all ACPSEM members. The working group is established and ready to move forward with these goals and is buoyed by the efforts of those who have previously completed DEI initiatives for the workforce.